\documentclass[conference]{IEEEtran}
\IEEEoverridecommandlockouts
\usepackage{cite}
\usepackage{amsmath,amssymb,amsfonts}
\usepackage{amsmath} 
\usepackage{algorithm} 
\usepackage{algpseudocode} 
\usepackage{graphicx}
\usepackage{textcomp}
\usepackage{xcolor}
\usepackage{tabularx}
\usepackage{booktabs} 
\usepackage{multirow} 

\usepackage{balance}
\usepackage{mathtools}
\usepackage{bm}
\usepackage[normalem]{ulem} 
\usepackage{hyperref}

\usepackage{courier}
\usepackage[utf8]{inputenc}
\usepackage{caption}
\usepackage{nicematrix}
\usepackage{fontawesome5}

\def\BibTeX{{\rm B\kern-.05em{\sc i\kern-.025em b}\kern-.08em
    T\kern-.1667em\lower.7ex\hbox{E}\kern-.125emX}}
\begin{document}

\title{Hierarchical Semantic Correlation-Aware Masked Autoencoder for Unsupervised Audio–Visual Representation Learning}

\author{\IEEEauthorblockN{Donghuo Zeng}
\IEEEauthorblockA{\textit{KDDI Research, Inc., Saitama, Japan} \\
do-zeng@kddi-research.jp}
\and
\IEEEauthorblockN{Hao Niu}
\IEEEauthorblockA{\textit{KDDI Research, Inc.,Saitama, Japan} \\
ha-niu@kddi.com}
\and
\IEEEauthorblockN{Masato Taya}
\IEEEauthorblockA{\textit{KDDI Research, Inc., Saitama, Japan} \\
ma-taya@kddi.com}
}


\maketitle
\begin{abstract}
Learning aligned multimodal embeddings from weakly paired, label-free corpora is challenging: pipelines often provide only pre-extracted features, clips contain multiple events, and spurious co-occurrences. We propose HSC-MAE (Hierarchical Semantic Correlation-aware Masked AutoEncoder), a dual-path teacher–student framework that enforces semantic consistency across three complementary levels of representation — from coarse to fine: (i) global-level canonical-geometry correlation via DCCA, which aligns audio and visual embeddings within a shared modality-invariant subspace; (ii) local-level neighborhood-semantics correlation via teacher-mined soft top-k affinities, which preserves multi-positive relational structure among semantically similar instances; and (iii)
sample-level conditional-sufficiency correlation via masked autoencoding, which ensures individual embeddings retain discriminative semantic content under partial observation. Concretely, a student MAE path is trained with masked feature reconstruction and affinity-weighted soft top-k InfoNCE; an EMA teacher operating on unmasked inputs via the CCA path supplies stable canonical geometry and soft positives. Learnable multi-task weights reconcile competing objectives, and an optional distillation loss transfers teacher geometry into the student. Experiments on AVE and VEGAS demonstrate substantial mAP improvements over strong unsupervised baselines, validating that HSC-MAE yields robust and well-structured audio–visual representations.
\end{abstract}


\section{Introduction}\label{sec:intro}
Learning aligned representations across heterogeneous modalities is a fundamental challenge in multimedia and embodied perception. Audio–visual representation learning seeks to discover shared semantic structure between sound and sight so that systems can ground concepts, reason about events, and act robustly in real-world scenes. Humans naturally fuse auditory and visual cues (e.g., a flash followed by thunder, the cadence of footsteps implying an approaching person) to form context-rich, resilient concepts; reproducing this ability in machines, without relying on costly manual labels, would enable perception and interaction in the wild. In this work, we focus on unsupervised approaches that learn cross-modal grounding from paired but unlabeled data, explicitly addressing the practical issues of noise, weak alignment, and limited access to raw sensor modalities.

Unsupervised audio–visual representation learning from paired but unlabeled clips is appealing but practically difficult. Real-world corpora are noisy and weakly aligned: clips often contain multiple concurrent events, temporal misalignment is common, and spurious co-occurrences violate the hard-positive assumption in standard contrastive learning. Two practical constraints further complicate the problem. First, many modern pipelines expose only compact, pre-extracted feature vectors (rather than raw pixels or spectrogram patches); methods must therefore build strong intra-modal semantics from these condensed descriptors. Second, cross-modal positives are inherently ambiguous, since many clips admit multiple valid positive matches across modalities. Consequently, the widely adopted one-positive-per-anchor assumption in contrastive learning misrepresents the local geometry of the embedding space and risks introducing confirmation bias~\cite{caron2020unsupervised}.

To address these issues, we introduce \textbf{HSC-MAE} (Hierarchical Semantic Correlation-Aware Masked Autoencoder), a dual-path teacher–student framework that explicitly enforces \emph{hierarchical semantic correlations} (HSC) across three complementary levels of representation — from coarse to fine: (1) \textit{Global-level canonical-geometry correlation} — audio and visual embeddings should share a common low-dimensional subspace that captures modality-invariant semantic factors. DCCA imposes this coarse distributional alignment, establishing a well-structured embedding manifold on which finer-grained constraints can build. (2) \textit{Local-level neighborhood-semantics correlation} — local neighborhoods should preserve multi-positive relational structure among semantically similar instances. Building on the global manifold, a soft top-k InfoNCE with teacher-mined affinities shapes these neighborhoods, relaxing the brittle single-positive assumption that misrepresents semantic relatedness in multi-event clips. (3) \textit{Sample-level conditional-sufficiency correlation} — individual embeddings must retain discriminative semantic content sufficient to predict missing feature dimensions. Sample-level masked autoencoding enforces this instance-level robustness, which is critical when working with pre-extracted descriptors where feature dimensions may be noisy or missing.

Together, these hierarchical constraints reduce a principled upper bound on retrieval risk. HSC-MAE realizes them with two coordinated training modes on a shared encoder: a \emph{student MAE path} (value-masked inputs) that enforces sample-level conditional sufficiency via reconstruction and local neighborhood consistency via soft top-k InfoNCE, and a \emph{CCA path} (clean inputs with gradient masking) that serves as Exponential Moving Average (EMA)~\cite{caron2021emerging} teacher enforcing global canonical geometry via DCCA. The teacher produces stable canonical embeddings and affinity weights that mine soft positives and provide geometric targets for optional distillation; the student learns corruption-robust local semantics while inheriting global structure from the teacher. An EMA teacher stabilizes neighborhood mining and prevents confirmation bias, while learnable multi-task weights reconcile competing objectives across the hierarchical levels during optimization.

We evaluate HSC-MAE on AVE and VEGAS benchmarks using mean Average Precision (mAP). Results show substantial and consistent gains over strong unsupervised baselines. Ablations confirm that global correlation, local neighborhood discrimination, and sample-level reconstruction each contribute complementary and non-redundant improvements to retrieval geometry. Contributions can be summarized as: (i) We introduce \textbf{HSC-MAE}, a dual-path masked autoencoder that enforces hierarchical semantic correlations — from global canonical geometry, through local neighborhood semantics, to sample-level conditional sufficiency — for unsupervised audio–visual representation learning from compact pre-extracted features. (ii) We introduce a teacher-guided soft top-k neighborhood mining strategy, where an EMA teacher produces stable affinity weights to define multi-positive relationships for contrastive learning. This formulation mitigates brittle one-positive assumptions and reduces confirmation bias in weakly aligned, multi-event settings.
(iii) We use a principled multi-task weighting scheme and optional distillation to reconcile reconstruction, correlation, and contrastive objectives, which stabilizes optimization across the hierarchical levels. (vi) We demonstrate significant mAP improvements on AVE and VEGAS and provide thorough ablations validating how the hierarchical constraints jointly and complementarily improve audio–visual cross-modal retrieval.

\section{Related Work}
\subsection{Audio–Visual Learning}
Classical methods align modalities with linear or shallow projections such as Canonical Correlation Analysis (CCA)~\cite{hardoon2004canonical}, while nonlinear extensions (DCCA) learn powerful shared subspaces via deep encoders~\cite{andrew2013deep}. More recent unsupervised methods leverage contrastive objectives adapted to paired audio–visual data to encourage cross-modal alignment~\cite{zeng2020deep, zeng2018audio,zeng2023learning, zeng2025metric}, and metric-learning variants such as triplet losses have also been applied to multimedia retrieval~\cite{schroff2015facenet}. These approaches, however, often assume strict positive pairings and can struggle on weakly aligned with multi-event clips. HSC-MAE builds on DCCA’s global alignment strengths while mitigating brittle pair assumptions via soft multi-positive contrast.
\vspace{-3pt}
\subsection{Masked Autoencoders}
Masked autoencoders (MAE) have shown strong self-supervision by reconstructing masked patches in images or spectrograms~\cite{he2022masked,huang2022masked}. When only pre-extracted feature vectors are available, patch masking is not applicable; instead, feature-level denoising and cross-modal prediction have been proposed (e.g., denoising autoencoders and feature reconstruction)~\cite{vincent2008extracting}. 
HSC-MAE adopts sample-level masking and feature-dimension reconstruction to learn robust intra-modal representations from compact features, and couples these reconstructions with cross-modal decoders so reconstruction benefits retrieval.
\vspace{-3pt}
\subsection{Metric Learning}
Metric learning losses (contrastive, triplet) provide many ways to shape embedding geometry~\cite{schroff2015facenet,hadsell2006dimensionality}. In unsupervised settings, mining reliable positives/negatives is critical: common strategies include top-$K$ neighbors, mutual nearest neighbors, clustering, and graph diffusion over kNN graphs~\cite{caron2020unsupervised}. 
Recent work advocates \emph{soft} positives or affinity-weighted losses to reflect uncertainty in mined labels~\cite{hoffmann2022ranking}. Our soft top-$k$ InfoNCE constructs weighted multi-positive targets from a teacher’s affinity estimates, reducing noise from ambiguous matches common in multi-event clips.
\vspace{-3pt}
\subsection{Knowledge distillation}
The exponential moving average (EMA) teacher models and bootstrapping methods (BYOL, DINO) have proven effective to avoid collapse and to provide stable targets in self-supervision~\cite{grill2020bootstrap,caron2021emerging}. In cross-modal learning, teacher-student schemes also stabilize mining and reduce confirmation bias by supplying smoother labels~\cite{zeng2025metric}. HSC-MAE leverages an EMA teacher both to produce stable correlation targets for soft neighborhood mining and to distill consistency into the student; combined with learnable multi-task weighting, this reduces optimization conflicts between reconstruction and alignment.

\section{Method}
\label{sec:method}
\subsection{Problem formulation and notation}
\label{sec:problem}
Let \(\mathcal{D}=\{(x_{a,i},x_{v,i})\}_{i=1}^N\) be an unlabeled collection of paired audio and visual feature vectors,
\(x_{a,i}\in\mathbb{R}^{d_a}\) and \(x_{v,i}\in\mathbb{R}^{d_v}\). Our goal is to learn modality encoders
\(f_a(\cdot;\theta)\) and \(f_v(\cdot;\theta)\) that map both modalities
into a shared embedding space \(\mathbb{R}^d\). Retrieval is performed by a similarity \(s(\cdot,\cdot)\) (e.g., cosine) on L2-normalized embeddings. HSC-MAE imposes semantic consistency across multiple levels of representation. 

\subsection{Dual-path forward pass}
\label{sec:dualpath}
HSC-MAE executes two coordinated forward modes on the same encoder parameters: (1) \textit{CCA-path (teacher/global geometry)}: Inputs are passed in a \emph{clean} mode (values preserved); gradients from selected input dimensions may be masked to protect canonical geometry. Clean-view embeddings \(Z_a^{\mathrm{cca}}, Z_v^{\mathrm{cca}}\) are optimized by a DCCA objective that encourages a shared low-dimensional subspace across modalities. An EMA (momentum) copy of the encoder is maintained as a teacher and evaluated on this CCA path; the teacher’s clean embeddings provide stable affinity estimates and geometric targets for mining and distillation. Gradients do not flow into the teacher. (2) \textit{MAE-path (student/robust semantics)}: Inputs are value-masked at the sample level (a fraction of feature dimensions zeroed) and passed through the same encoder and a decoder to reconstruct missing components. Masked-view embeddings \(Z_a^{\mathrm{mae}}, Z_v^{\mathrm{mae}}\) are trained with reconstruction, a teacher-guided soft multi-positive InfoNCE, and an optional consistency loss that aligns student embeddings to teacher geometry. This path encourages embeddings that are locally coherent and robust to partial observations while inheriting global structure from the teacher.

\begin{figure}[t]
  \centering
  \includegraphics[width=0.92\linewidth]{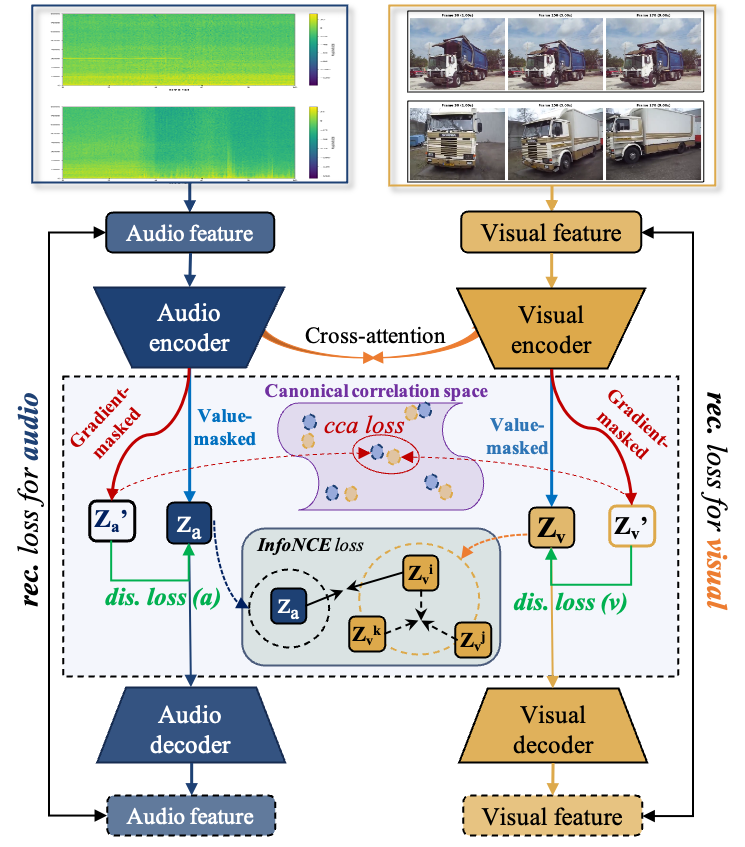}
  \caption{\textbf{Overview of the HSC-MAE architecture.} Pre-extracted audio and visual features are processed by shared encoders and a cross-attention fusion block under two coordinated training modes. 
    The \emph{student MAE path} applies sample-level value masking and is optimized by reconstruction, a teacher-guided soft top-$k$ InfoNCE, producing robust embeddings $(Z_a, Z_v)$. 
    In parallel, the \emph{EMA teacher CCA path} preserves input values and enforces global cross-modal alignment via DCCA, yielding clean embeddings $(Z'_a, Z'_v)$. 
    The teacher provides stable semantic affinities and geometric targets for neighborhood mining and distillation, while gradients are blocked from flowing into the teacher.}
  \label{fig:arch}
\end{figure}

\subsection{HSC-MAE architecture}
\label{sec:architecture}
The architecture (Fig.~\ref{fig:arch}) enforces hierarchical semantic correlations across three complementary levels — global subspace, local neighborhood, and individual sample — via four objectives detailed below. Consistent with the hierarchy introduced in Sec.~\ref{sec:intro}, we present them from coarse to fine.

\subsubsection{Global shared-subspace alignment (canonical-geometry correlation)}
Denote the clean-view audio-visual embeddings by \(Z_a^{\mathrm{cca}},Z_v^{\mathrm{cca}}\). DCCA finds canonical directions whose correlations \(\{\rho_m\}_{m=1}^r\) are maximized. We minimize the negative sum of top-\(r\) canonical correlations:
\[
\mathcal{L}_{\mathrm{cca}} \;=\; -\sum_{m=1}^{r} \rho_m\big(Z_a^{\mathrm{cca}},Z_v^{\mathrm{cca}}\big).
\]
Maximizing these correlations encourages audio and visual representations to occupy a common low-dimensional subspace of modality-invariant factors. It establishes the global geometric structure on which local and sample-level constraints are built.

\subsubsection{Local neighborhood preservation (neighborhood-semantics correlation)}

Let \(Z_i^{\mathrm{mae}}\) be the masked-view embedding of sample \(i\) and let \(W_{ij}\ge 0\) be nonnegative affinity weights produced by the teacher over the mined top-k neighborhood, normalized so that \(\sum_j W_{ij}=1\). Using cosine similarity \(s(\cdot,\cdot)\) and temperature \(\tau\), we define:
\[
\mathcal{L}_{\mathrm{infoNCE}}
= -\mathbb{E}_{i}\Big[\sum_{j} W_{ij}\log\frac{\exp\big(s(Z_i^{\mathrm{mae}},Z_j^{\mathrm{mae}})/\tau\big)}
{\sum_{\ell}\exp\big(s(Z_i^{\mathrm{mae}},Z_\ell^{\mathrm{mae}})/\tau\big)}\Big].
\]
This affinity-weighted multi-positive objective shapes local neighborhood structure within the globally aligned manifold, relaxing the brittle single-positive assumption by tolerating multiple plausible positives per anchor — a necessary property for weakly aligned, multi-event clips.

\subsubsection{Sample-level MAE reconstruction (conditional sufficiency correlation)}

Given an input $x \sim \mathcal{D}$, a masked view $\tilde{x} \sim q(\tilde{x} \mid x)$ is generated via a stochastic masking process, and reconstructed as $\hat{x}$=$D(E(\tilde{x}))$. The reconstruction objective minimizes the mean squared error:
\[
\mathcal{L}_{\mathrm{rec}} 
= \mathbb{E}_{x \sim \mathcal{D},\, \tilde{x} \sim q(\tilde{x}\mid x)} 
\big[\|x - \hat{x}\|_2^2\big].
\]
Minimizing $\mathcal{L}_{\mathrm{rec}}$ requires the masked-view embedding $Z^{\mathrm{mae}}$=$E(\tilde{x})$ to retain sufficient statistics of $x$ given the corrupted observation $\tilde{x}$, thereby enforcing conditional sufficiency at the sample level and ensuring robustness to noisy or incomplete descriptors common in pre-extracted feature pipelines.

\paragraph{Optional consistency (distillation)}
The EMA teacher evaluated on the clean path produces stable teacher embeddings \(Z^{t}\). We optionally align student embeddings to these targets:
\[
\mathcal{L}_{\mathrm{dis}} \;=\; \mathbb{E}_i\big[\|Z_i^{\mathrm{mae}}-Z_i^{t}\|_2^2\big].
\]
This distillation transfers teacher geometry into the masked student and reduces confirmation bias during mining.

\paragraph{Overall objective and optimization}
The individual terms are combined with learnable log-variance weights
\(\boldsymbol{\sigma}=(\sigma_{\mathrm{rec}},\sigma_{\mathrm{infoNCE}},\sigma_{\mathrm{cca}},\sigma_{\mathrm{dis}})\)
following an uncertainty-based scheme~\cite{kendall2018multi}:
\[
\mathcal{L}_{\mathrm{total}}(\theta,\boldsymbol{\sigma})
=\sum_{m}\exp(-\sigma_m)\,\mathcal{L}_m(\theta)+\sigma_m,
\]
where the sum runs over the active losses \(m\in\{\mathrm{rec,infoNCE,cca,dis}\}\).
Both encoder parameters \(\theta\) and the log-variance parameters \(\boldsymbol{\sigma}\) are updated by gradient descent. The EMA teacher parameters \(\theta^t\) are updated with a momentum rule
\(\theta^t\leftarrow \rho\theta^t + (1-\rho)\theta\) after each optimization step; gradients are not propagated into \(\theta^t\). Motivated by DCCA~\cite{andrew2013deep}, we append a linear CCA projection layer at the end of the encoder network, applied during both training and inference. This layer continuously reduces embedding dimensionality and emphasizes maximally correlated canonical directions, yielding a more discriminative and geometrically structured retrieval space throughout optimization.
\section{Experiments}
\label{sec:experiments} 
\subsection{Datasets and Evaluation}
We evaluate HSC-MAE on two standard audio–visual benchmarks~\cite{zeng2023learning}: AVE~\cite{zeng2023learning}, consisting of 1,955 clips across 15 categories (1,564 for training and 391 for testing), and VEGAS~\cite{zeng2023learning}, containing 28,103 YouTube clips (2–10s) from 10 categories (22,482 for training and 5,621 for testing). Audio features are extracted as 128-D embeddings using VGGish~\cite{hershey2017cnn}, while visual features are 1,024-D descriptors derived by average-pooling InceptionV3~\cite{abu2016youtube} frame-level features. Unsupervised cross-modal retrieval (UCMR) is evaluated in both audio-to-visual (A2V) and visual-to-audio (V2A) directions using cosine similarity between embeddings. The final score is the mean of the two Mean Average Precision (MAP) values, following prior works~\cite{zeng2023learning, zeng2022complete}. Evaluation is class-based, using dataset categories only for testing, not for training.

\subsection{Implementation Details}
HSC-MAE employs modality-specific three-layer MLP encoders. The \textit{audio encoder} ([128,1024,1024,1024]) and \textit{visual encoder} ([1024,1024,1024,1024]) use BatchNorm in the first layer, LayerNorm thereafter, \textit{Tanh} as activations, and Dropout (0.2). 
Cross-modal fusion is achieved via multi-head attention (64 heads).
Linear projectors map audio and visual embeddings to a 32-D retrieval space. Training uses a dual-path masking scheme: (i) GradMask in the CCA path to enable selective gradient flow, and (ii) value masking in the MAE path for reconstruction.
Mask ratios are 0.3 for VEGAS and 0.2 for AVE. Soft top-$k$ mining ($k$=5) selects pseudo-positive pairs via model predictions, selections are weighted via a temperature-scaled softmax ($\tau$=0.05) and used in symmetric InfoNCE losses. Loss terms are balanced with uncertainty-based multi-task weighting, preceded by a 5-epoch warmup with fixed weights ([$\mathcal{L}_{\mathrm{rec}}, \mathcal{L}_{\mathrm{cca}}, \mathcal{L}_{\mathrm{dis}}, \mathcal{L}_{\mathrm{infoNCE}}$] = $[1,\ \text{epoch}{\times}0.1,\ 0.1,\ 0.05]$) to ensure stable initialization before learnable balancing is introduced. This warmup prevents cold-start sensitivity and yields consistent weight convergence across runs without manual tuning.

We optimize with AdamW (LR=3$\times$10$^{-4}$, weight decay = 1$\times$10$^{-4}$), gradient clipping = 1.0, a cosine-annealing schedule with $T_{max}$ = 50. A teacher network is maintained via EMA of student weights, with momentum $\rho$ annealed from 0.95 to 0.999. After training, a linear CCA (output dim=10) is fitted on the training embeddings for final retrieval alignment. Batch sizes are 512 (VEGAS) and 400 (AVE); models are trained for 100 epochs. All models were trained on an NVIDIA RTX 3080 GPU (10 GB). Code available at~\url{https://github.com/ZenzenDatabase/UnsupervisedMAE}

\subsection{Baselines}
We evaluate HSC-MAE against a diverse set of classical and recent \emph{unsupervised} audio--visual retrieval baselines. All methods use the same pre-extracted features and, where applicable, are reimplemented with comparable encoder capacity and training budgets to ensure fair comparison. Specifically, we include a \textit{Random} baseline as a lower-bound reference. Classical correlation-based methods include \textit{linear CCA}~\cite{hardoon2004canonical}, \textit{kernel CCA (KCCA)}~\cite{akaho2006kernel}, and \textit{deep CCA (DCCA)}~\cite{andrew2013deep} implemented with shallow MLP encoders. We further compare against \textit{contrastive learning} objectives, including \textit{InfoNCE with single positives}~\cite{oord2018infonce} using same-sample audio--visual pairs, a \textit{CLIP-style symmetric cross-modal contrastive loss}~\cite{radford2021clip}, and \textit{triplet-based metric learning}~\cite{schroff2015facenet}. Finally, we benchmark against recent unsupervised cross-modal hashing and masked autoencoding approaches, including \textit{DECH}~\cite{yang2023dech}, \textit{DUMCH}~\cite{wang2022dumch}, \textit{UCCH}~\cite{hu2022unsupervised}, and \textit{CAV-MAE}~\cite{gong2022contrastive}. Whenever feasible, baselines are retrained using the identical training settings. We report results of CCA and KCCA from the original publications.

\begin{table}[t]
\small
\centering
\caption{\textbf{mAP comparison across state-of-the-art methods.} The best results are shown in bold, and the second-best results are underlined.}
\setlength{\tabcolsep}{5pt}
\begin{NiceTabular}{l|ccc|ccc}[cell-space-limits=0.5pt]
\toprule
\Block{2-1}{\textbf{Methods}} & \Block{1-3}{\textbf{AVE dataset}} & &  & \Block{1-3}{\textbf{VEGAS dataset}} & & \\ 
\cline{2-7}
& A2V & V2A & Avg.
& A2V & V2A & Avg. \\ 
\hline
Random
& 0.127 & 0.124 & 0.126
& 0.110 & 0.109 & 0.109\\
CCA
& 0.190 & 0.189 & 0.190
& 0.332 & 0.327 & 0.330\\
KCCA
&0.3062 &0.1818 &0.2440
& 0.288 & 0.273 & 0.281\\
DCCA
& 0.3382 & 0.3438 & 0.3410
& \underline{0.7541} & 0.7574 & \underline{0.7558}\\
InfoNCE
&0.4646 &0.4643 &0.4644
&0.7359 &0.7097 &0.7228 \\
Contrastive
&0.3943 &0.3912 &0.3927
& 0.5089 &0.5101 &0.5095 \\
Triplet
&0.4519 &0.4524 &0.4521
&0.4306 &0.4310 &0.4308 \\
DUMCH 
& 0.4611 &0.4719 &0.4665
&0.5600 &0.5439 &0.5519\\
UCCH
& 0.4535 & 0.4498 & 0.4517
& 0.5986 & 0.6004 & 0.5995 \\
DECH
& 0.3368  &0.3346  &0.3357 
&0.6280 &0.6257 &0.6269\\
CAV-MAE
& \underline{0.6123} & \underline{0.6207} & \underline{0.6165} 
& 0.7453 &\underline{0.7617} &0.7535 \\
\hline
\textbf{Ours}
& \textbf{0.7747} & \textbf{0.7728} & \textbf{0.7737}
& \textbf{0.8013} & \textbf{0.8039} & \textbf{0.8026}\\
\bottomrule
\end{NiceTabular}
\label{table:comparison}
\end{table}

\begin{table*}[t]
\small
\centering
\caption{Component-wise impact on the final objective loss for different methods on the AVE and VEGAS datasets.}
\label{tab:ablation}
\renewcommand{\arraystretch}{1.3}
\begin{NiceTabular}{*{4}{wc{1.2cm}}|ccc|ccc}[cell-space-limits=3pt]
\toprule
\Block{2-1}{\textbf{CCA}} & \Block{2-1}{\textbf{Rec.}} & \Block{2-1}{\textbf{Soft InfoNCE}} & \Block{2-1}{\textbf{Self-distillation}} & \Block{1-3}{\textbf{AVE dataset}} & & & \Block{1-3}{\textbf{VEGAS dataset}} & & \\ 
\cline{5-10}
& & & & A2V & V2A & Avg. & A2V & V2A & Avg. \\ 
\midrule 
\checkmark & \checkmark & \checkmark & \checkmark & \textbf{0.7747} & \textbf{0.7728} & \textbf{0.7737} & \textbf{0.8084} & \textbf{0.8228} & \textbf{0.8156} \\
$-$ & \checkmark & \checkmark & \checkmark & 0.6229 & 0.6239 & 0.6234 & 0.7750 & 0.7882 & 0.7816 \\
\checkmark & \checkmark & \checkmark & $-$ & 0.7486 & 0.7427 & 0.7457 & 0.8060 & 0.8102 & 0.8081 \\
\checkmark & \checkmark & $-$ & \checkmark & 0.6735 & 0.6746 & 0.6741 & 0.7805 & 0.7749 & 0.7777 \\
\checkmark & \checkmark & $-$ & $-$ & 0.5728 & 0.5701 & 0.5715 & 0.7903 & 0.7894 & 0.7899 \\
\checkmark & $-$ & $-$ & \checkmark & 0.6143 & 0.6145 & 0.6144 & 0.7940 & 0.7883 & 0.7911 \\
\checkmark & $-$ & $-$ & $-$ & 0.5688 & 0.5698 & 0.5693 & 0.7891 & 0.7856 & 0.7873 \\
\bottomrule 
\end{NiceTabular}
\end{table*}

\subsection{Main results}
Table~\ref{table:comparison} reports retrieval performance (mAP) of HSC-MAE and all baselines on AVE and VEGAS. HSC-MAE consistently outperforms classical correlation-based methods, contrastive objectives, and recent MAE-based approaches across both retrieval directions (A2V and V2A) as well as averaged mAP.

Quantitatively, HSC-MAE substantially improves upon the strongest prior method, CAV-MAE, on AVE. Specifically, A2V mAP increases from 0.6123 to 0.7747 (+26.24\%), V2A from 0.6207 to 0.7728 (+15.21\%), and the averaged mAP from 0.6165 to 0.7737 (+15.72\%). On VEGAS, the improvements are smaller but consistent: A2V improves from 0.7453 to 0.8013 (+15.60\%), V2A from 0.7617 to 0.8039 (+4.22\%), and the averaged mAP from 0.7535 to 0.8026 (+4.91\%). Compared with standard contrastive learning baselines, including Contrastive, InfoNCE, and Triplet losses, HSC-MAE achieves markedly higher performance on both datasets, demonstrating that hierarchical semantic modeling and the proposed dual-path masking strategy provide more effective cross-modal alignment than vanilla contrastive objectives.

\subsection{Ablation studies}
We conduct ablation studies to quantify the impact of individual loss components and mask ratio in HSC-MAE, validating the design choices described in Sec.~\ref{sec:method}.

\subsubsection{Impact of loss components}
We evaluate the contribution of each major component in HSC-MAE by selectively removing or modifying individual components. The ablated variants include removing self-distillation (w/o EMA), disabling the CCA objective for global alignment, replacing soft top-\(k\) mining with strict single-positive InfoNCE, and etc. The quantitative impact of these ablations on AVE and VEGAS is summarized in Table~\ref{tab:ablation}. Removing any component leads to a consistent performance drop, with the largest degradation observed when masked reconstruction or soft InfoNCE is removed, confirming their central role in learning robust cross-modal representations. Disabling CCA further degrades performance, highlighting the importance of preserving global cross-modal correlation.

\begin{figure}[h]
  \centering
  \includegraphics[width=0.95\linewidth]{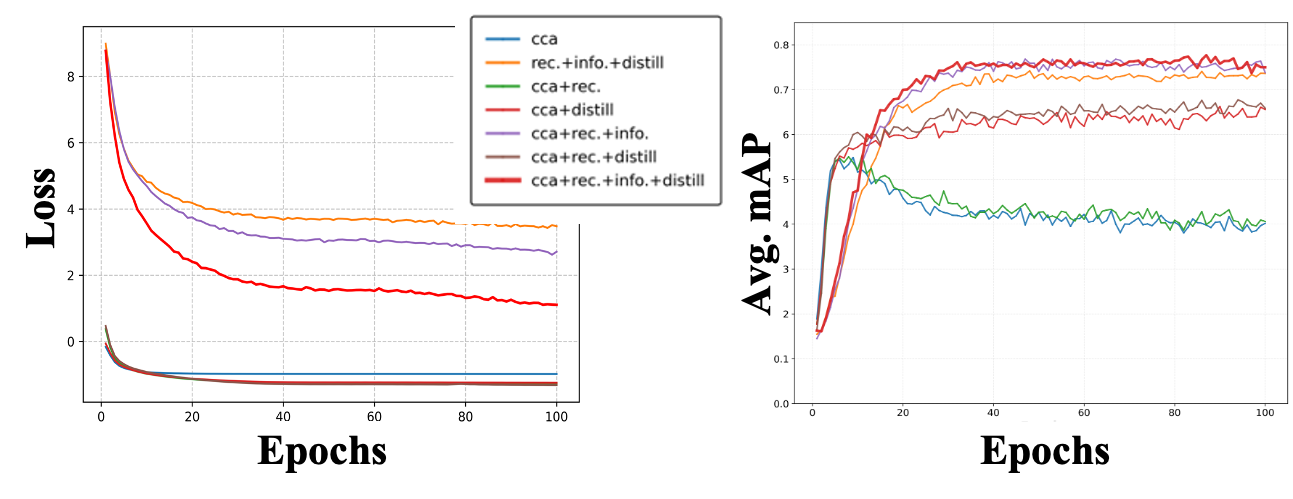}
  \caption{(Left) Decomposition of training losses; (Right) test mAP over epochs (1-100) on AVE under component-wise ablations.}
  \label{fig:loss_map}
\end{figure}

Figure~\ref{fig:loss_map} visualizes the training loss decomposition and corresponding test mAP under different ablations. Removing self-distillation or CCA leads to slower convergence and higher final losses, accompanied by degraded retrieval accuracy, whereas omitting InfoNCE yields fast, stable convergence with low loss but inferior mAP, indicating weak representation geometry and limited generative capacity due to missing local geometric constraints. In contrast, the full model exhibits smoother optimization and consistently higher mAP, validating the effectiveness of HSC-MAE.

\begin{figure}[h]
  \centering
  \includegraphics[width=0.9\linewidth]{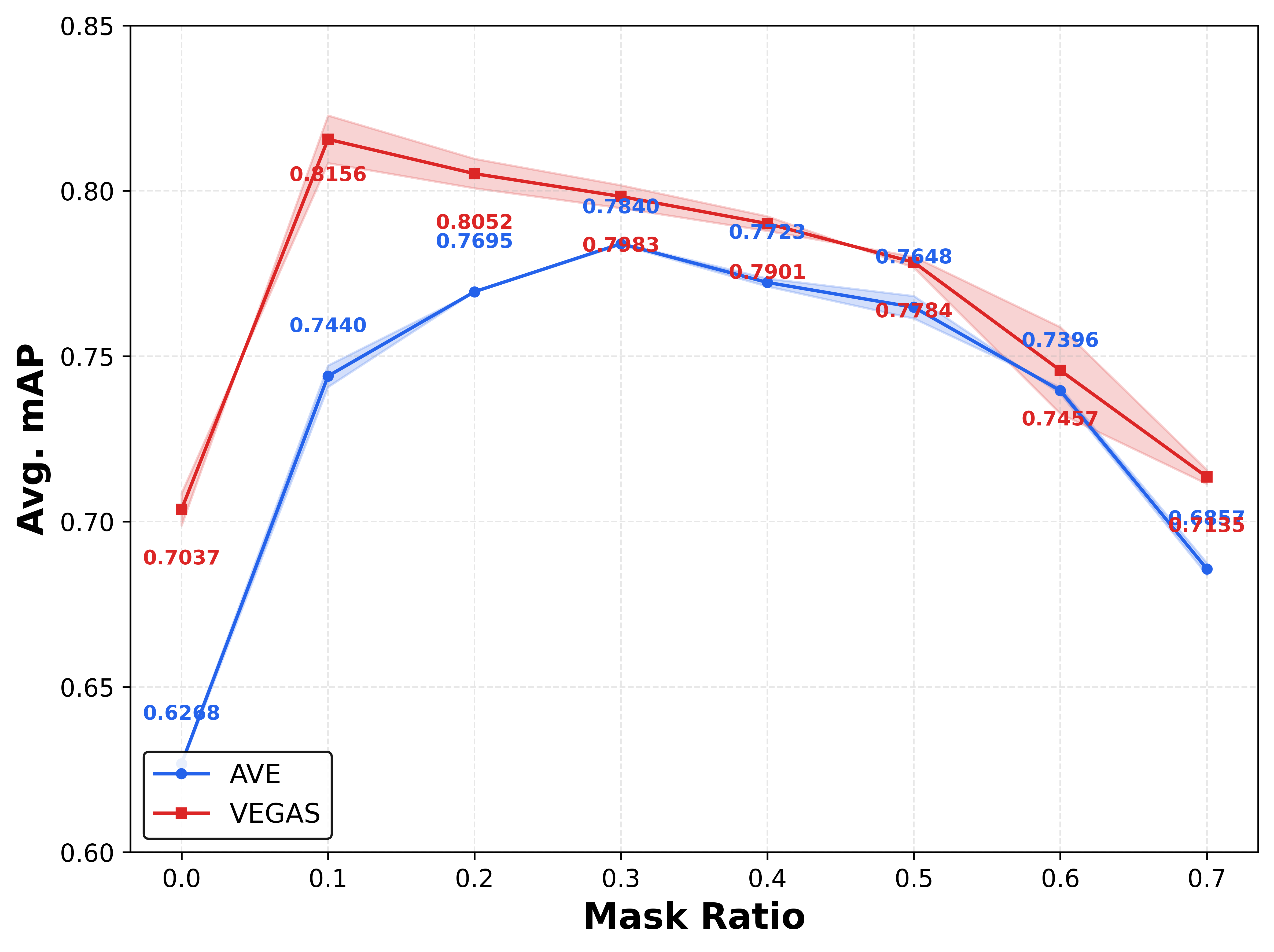}
     \caption{Effect of mask ratio on UCMR task on both AVE and VEGAS datasets.  Solid curves show the average mAP, while shaded regions indicate the absolute gap between the two retrieval directions.
    }
  \label{fig:mask_ratio}
\end{figure}

\subsubsection{Impact of mask ratio}
We analyze the sensitivity of HSC-MAE to the masking ratio, evaluated over \(\{0.0, 0.1, 0.2, 0.3, 0.4, 0.5, 0.6, 0.7\}\) on both AVE and VEGAS. Figure~\ref{fig:mask_ratio} illustrates the effect of masking on retrieval performance (mAP). For both datasets, performance improves rapidly as the mask ratio increases from 0.0, peaks at moderate values (0.1 and 0.3), and degrades when masking becomes too aggressive. This behavior highlights the trade-off between learning robust representations through sample-level masked reconstruction and preserving sufficient semantic information for cross-modal alignment.

We additionally observe dataset-dependent effects of the mask ratio in the shaded regions of Figure~\ref{fig:mask_ratio}: on the larger VEGAS, moderate masking (0.2, 0.3) yields the highest mAP but a larger A2V–V2A gap, indicating residual directional bias, whereas on the smaller AVE dataset the gap is consistently smaller, reflecting more symmetric embeddings; in both cases, low masking under-regularizes the model and excessive masking disrupts cross-modal semantics.
\begin{figure}[h]
\centering
\includegraphics[width=0.45\textwidth]{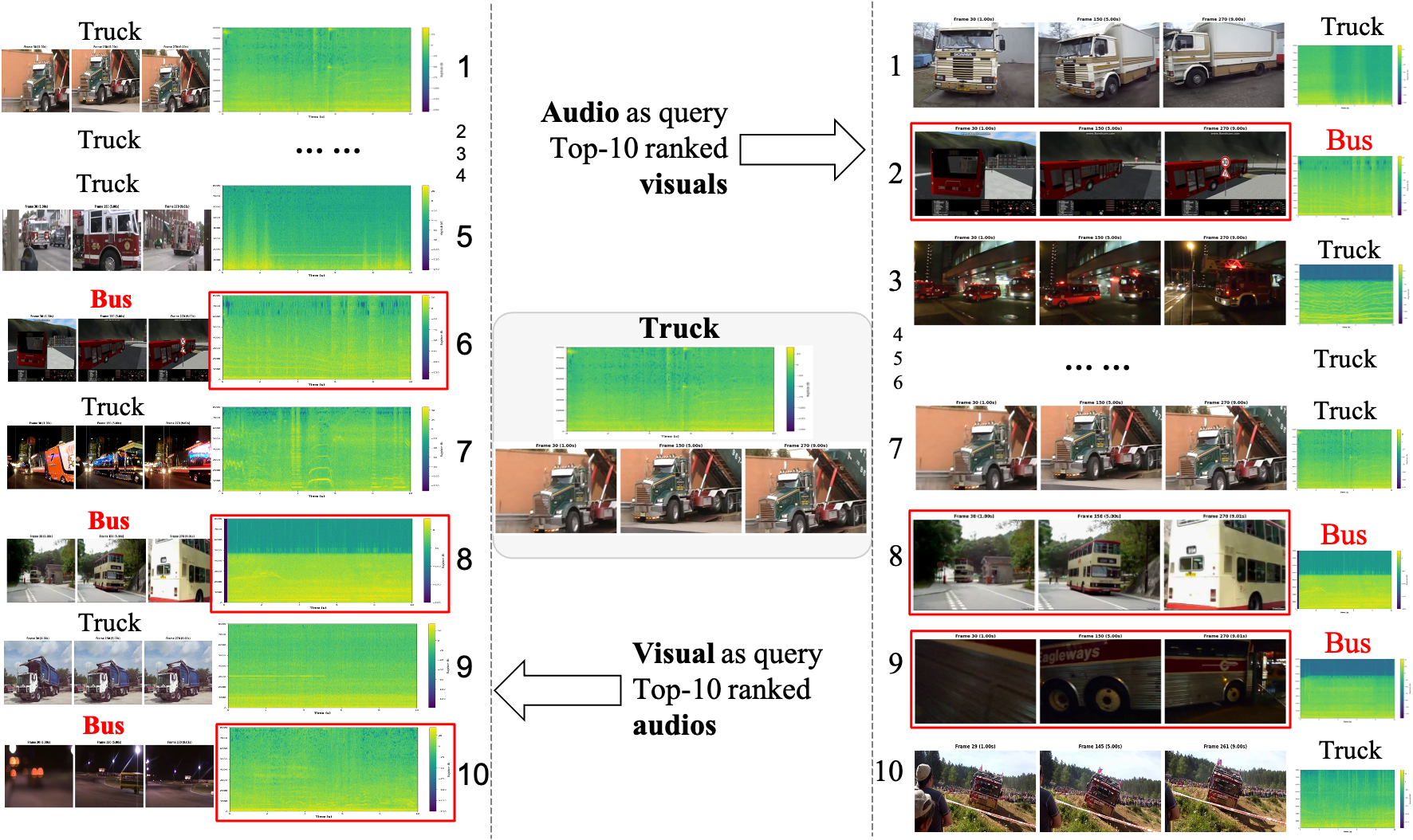}
\caption{Qualitative audio--visual cross-modal retrieval results on AVE. For each query (audio or visual), the top-10 retrieved results are shown.}
\label{fig:retrieval_case}
\end{figure}
\vspace{-5pt}
\subsection{Qualitative Results}
Figure~\ref{fig:retrieval_case} shows A2V and V2A retrieval examples on AVE to illustrate semantic alignment. Using a \emph{truck} audio as the A2V query, the model returns 7 correct \emph{truck} visuals in the top-10 (non-matches: \emph{bus} at ranks 2, 8, 9). Using the paired \emph{truck} visual for V2A, retrieval again yields 7 correct audio clips with \emph{bus} confusions at ranks 6, 8, 10. These results indicate HSC-MAE reliably captures cross-modal semantics; remaining errors are primarily between acoustically and visually similar classes (e.g., \emph{bus} vs.\ \emph{truck}), reflecting inherent category ambiguity rather than misalignment.

\section{Conclusion}
We introduced HSC-MAE that combines sample-level masked reconstruction, DCCA-driven global alignment, and teacher-guided soft top-$k$ contrast within a dual-path teacher–student design. By decoupling reconstruction and correlation objectives—and coupling them through cross-attention, an EMA teacher, and learnable multi-task weights, HSC-MAE enforces semantic consistency at multiple structural levels: canonical geometry (shared modality subspace), neighborhood semantics (multi-positive local structure), and conditional sufficiency (robust intra-modal features). Experiment shows substantial mAP improvements on AVE and VEGAS compared to unsupervised baselines. Ablations confirm that reconstruction, soft-neighborhood discrimination, and global correlation each provide complementary benefits, and their combination yields the best retrieval geometry.

Beyond retrieval, HSC-MAE offers a practical approach for label-scarce multimodal grounding in embodied agents and related systems. Future work will explore additional modalities, finer temporal modeling, and scalability improvements for large-scale or online settings.

\bibliographystyle{IEEEbib}
\bibliography{refs}

\end{document}